\documentclass[superscriptaddress]{revtex4-1}
\usepackage{graphicx,ulem}  % needed for figures
\usepackage{dcolumn}   % needed for some tables
\usepackage{bm}        % for math
\usepackage{amssymb}   % for math
\usepackage{amsmath}
\usepackage{blkarray}
\usepackage{textcomp}
\usepackage{color}
\usepackage{flexisym}
\usepackage{sidecap}
\usepackage{booktabs}
\usepackage{mathtools}

\makeatletter
\let\cat@comma@active\@empty
\makeatother

\DeclareMathAlphabet{\mathpzc}{OT1}{pzc}{m}{it}

\begin{document}
%\linenumbers
\author{Sulimon Sattari}
\affiliation{Research Center of Mathematics for Social Creativity, Research Institute for Electronic Science, Hokkaido University, Kita 20, Nishi 10, Kita-ku, Sapporo 001-0020, Japan}

\author{Udoy~S.~Basak}
\affiliation{Graduate School of Life Science, Transdisciplinary Life Science Course, Hokkaido University, Kita 12, Nishi 6, Kita-ku, Sapporo 060-0812, Japan}
\affiliation{Pabna University of Science and Technology, Pabna 6600, Bangladesh}%

\author{Ryan G. James}
\affiliation{Department of Physics, Complexity Sciences Center, University of California, Davis, CA 95616, USA}

\author{James P. Crutchfield}
\affiliation{Department of Physics, Complexity Sciences Center, University of California, Davis, CA 95616, USA}

\author{Tamiki Komatsuzaki}
\affiliation{Graduate School of Life Science, Transdisciplinary Life Science Course, Hokkaido University, Kita 12, Nishi 6, Kita-ku, Sapporo 060-0812, Japan}
\affiliation{Research Center of Mathematics for Social Creativity, Research Institute for Electronic Science, Hokkaido University, Kita 20, Nishi 10, Kita-ku, Sapporo 001-0020, Japan}
\affiliation{Institute for Chemical Reaction Design and Discovery (WPI-ICReDD), Hokkaido University Kita 21 Nishi 10, Kita-ku, Sapporo, Hokkaido 001-0021, Japan}

\affiliation{Graduate School of Chemical Sciences and Engineering Materials Chemistry and Engineering Course, Hokkaido University, Kita 13, Nishi 8, Kita-ku, Sapporo 060-0812, Japan}

\title{Modes of information flow in collective cohesion}

\date{\today}
\begin{abstract}
Pairwise interactions between individuals are taken as fundamental drivers of
collective behavior---responsible for group cohesion and decision-making. While
an individual directly influences only a few neighbors, over time indirect
influences penetrate a much larger group. The abiding question is how this
spread of influence comes to affect the collective. In this, one or a few
individuals are often identified as ``leaders'', being more influential than
others in determining group behaviors. To support these observations transfer
entropy and time-delayed mutual information are used to quantitatively identify
underlying asymmetric interactions, such as leader-follower classification in
aggregated individuals---cells, birds, fish, and animals. However, these
informational measures do not properly characterize asymmetric interactions.
They also conflate distinct functional modes of information flow between
individuals and between individuals and the collective. Computing information measures 
conditioning on multiple agents requires the proper sampling
of a probability distribution whose dimension grows exponentially with the number of agents being conditioned on. 
This is not feasible in practice. Employing simple models
of interacting self-propelled particles, we examine the pitfalls of using
time-delayed mutual information and transfer entropy to quantify 
the strength of influence from a leader to a follower.
Surprisingly, one must be wary of these pitfalls even for two interacting
particles. As an alternative we decompose transfer entropy and time-delayed
mutual information into intrinsic, shared, and synergistic modes of information
flow. The result not only properly reveals the underlying effective
interactions, but also facilitates a more detailed diagnosis of how individual
interactions lead to collective behavior. This exposes, for example, the role
of individual and group memory in collective behaviors. In addition,
we demonstrate in a multi-agent system how knowledge of the decomposed information modes between a single pair of agents
reveals the nature of many-body interactions without conditioning on additional agents.

\end{abstract}
\maketitle
\section{Introduction}

Coherent collective behavior fascinates us when a global pattern emerges from
individuals who share information with others only in their local vicinity.
Decisions made by one individual apparently cascade throughout the entire
group. That said, not all members have the same influence. The challenge here
in explaining such emergent behaviors is to infer the underlying relationships
among individuals from observations. Not surprisingly, this challenge has
attracted many over decades to diagnose collective behaviors in a variety of
systems
\cite{Copenhagen2016,Hornischer2019,Nagy2010,Porfiri2018,Butail2016,Torney2018,Strandburg2015}.
In epithelial Madin-Darby canine kidney monolayers, for example, collective
cell migration is triggered by multicellular protrusions forming a fingerlike
structure \cite{Reffay2011,Reffay2014,Yamaguchi2015}. Photoablating a group of
cells from the finger tip makes the remaining cell group lose its sense of
direction. The interpretation is that the former and latter cells have acted as
if they were  ``leaders'' and ``followers''. This functional assignment can be
carried out due to their spatial location along the protrusion and since the
two cell classes are genetically distinct~\cite{Yamaguchi2015}.

% R Mayor, S Etienne-Manneville - Nature reviews Molecular cell biology, 2016
%Interplay of RhoA and mechanical forces in collective cell migration driven by leader cells
%M. Reffay, M. C. Parrini, O. Cochet-Escartin, B. Ladoux, A. Buguin, S. Coscoy, F. Amblard, J. Camonis & P. Silberzan
%Nature Cell Biology volume 16, pages217–223(2014)

Identifying leaders---even defining what that role means~\cite{Bollt2018}---is
very difficult. Especially so, when probing the mechanisms that cause the
dynamical behaviors of aggregated agents. Beyond cells, these basic questions
also apply to bird flocking~\cite{Nagy2010}, fish
schooling~\cite{Porfiri2018,Butail2016}, caribou migration~\cite{Torney2018},
and baboon foraging~\cite{Strandburg2015}. A leader agent is often defined
as an individual that exerts more influence upon others than others influence
on upon it. That is, the role is fundamentally asymmetric. Previous
studies~\cite{Basak2020,Porfiri2018,Butail2016,Mwaffo2017,Mwaffo2018,Mwaffo2018_2,Takamizawa2019}
proposed that, under this definition, pairwise analysis of trajectories can
assign leaders and followers under the working hypothesis that a change in
motion of the leader forecasts a change in motion of the follower. From
this, one interprets the change of leader motion as a candidate cause
that triggers the motion of followers.

Various statistical quantities are used to infer causal relationships~\cite{Liu2021}. 
In pigeon flocks, for example, time-delayed correlation between
the orientation of individuals at one time instance and the orientation of
others at previous times reveals a hierarchical leadership structure and also
provides a method to quantify the timescale of influence~\cite{Nagy2010}. In
such a case, the motion of one pigeon is correlated with the past motion of
another. Granger causality~\cite{Bressler2011} is seen as an improvement to
time-delayed correlation as it quantifies the predictability of the current
state of a variable based on knowledge of a variable at a previous time.
Time-delayed correlation and Granger causality both assume linear relationships
between variables, though. This generally does not hold. More recent studies
argued that information-theoretic quantities---transfer entropy, time-delayed
mutual information, and causation entropy---are superior when quantifying influence
since they naturally accommodate the highly nonlinear nature of
multi-agent systems~\cite{Ramirez2021,Basak2020,Basak2021,Butail2016,Mwaffo2017,Mwaffo2018,Orange2015,Lord2016,Jeong2001,Sun2015,Lord2016,Brown2020,Barnett2013}.

In practice, one must consider 
the potential for mis-classifying influence when using information-theoretic methods. 
Carefully considering the definitions of 
information-theoretic quantities---such as, transfer entropy
or time-delayed mutual information---further illuminates the types of influence 
a particular individual has. As pointed out in introducing transfer entropy
\cite{Schreiber2000}, time-delayed mutual information reports a nonzero value
between the present of a stochastic variable $X$ and the future of a stochastic
variable $Y$ even when $Y$ has no direct influence on $X$. This implies that it
cannot be directly employed to infer the underlying mutual influence among
individuals. It also includes additional information not intrinsically coming
from $X$.

Transfer entropy, in contrast, computes the reduction of uncertainty about
$Y$'s future while knowing $X$'s present, conditioned on $Y$'s present.
Recently though, Ref.~\cite{James2018} showed that, paralleling time-delayed
mutual information, transfer entropy incorporates additional, unwarranted
information; namely, the reduction of uncertainty about $Y$ that occurs by
knowing the present state of $X$ and $Y$ simultaneously. This information is
extraneous to determining ``flow'' and, misleadingly, adds to the desired
information: intrinsic flow from $X$ to $Y$. In this view, transfer entropy
decomposes into two distinct modes of information flow---intrinsic and
synergistic~\cite{James2018}.

Pairwise interactions are fundamental to information theory's development of
input-output (``two-port'') communication channels \cite{Cove06a}. As such,
they provide a primary statistical tool that, as we show, makes it possible to
infer the underlying influences among individuals. To obtain maximum insight
into the mechanisms underlying multi-agent systems, the following focuses on
decomposing transfer entropy and time-delayed mutual information into Ref.
\cite{James2018}'s three different fundamental modes of information
flow---termed \emph{intrinsic}, \emph{shared}, and \emph{synergistic}
information flows. The results demonstrate how the decomposed elemental
information flows shed light on the influences that drive leader-follower
relationships.

As an illustrative vehicle we employ a generalized Vicsek
model~\cite{Vicsek1995} with two additional features: 1) tunable influence
weight of one particle over another (i.e., leaders have larger influence) and
2) particle memory. We show that, by analyzing the effects of 1) and 2) on the
three modes of information flow, intrinsic information flow exists whenever the
motion of an agent depends on another with nonzero weight, as does transfer
entropy and time-delayed mutual information. Shared and synergistic
information, however, can occur when agents mutually influence each other and
synergistic information only occurs in such cases. These results extend
previous studies on modes of information flow in finite-state hidden Markov
models by introducing Vicsek models that are fundamental to understanding
collective motion and exhibit distinct modes of information flow. Moreover, we
probe the effect of agent memory and the effect of more than two interacting
agents on the different modes of information flow and their role in collective
behavior.

\section{Background: Measuring causal influence}

We now review several information-theoretic measures of statistical
interdependence---measures that have been offered up as ways to detect causal
influence. With these in hand, we turn to explore how useful (or not) they
are in analyzing leader-follower relationships.

\subsection{Detecting causal influence via time-delayed mutual information and transfer entropy}
\label{sec: inference_TE}

Our definition says that leaders are, on average, more influential than
followers. As a consequence of this asymmetry, a follower\textquotesingle s
behavior is affected by the leader\textquotesingle s, but there is a time
delay. To determine the degree of causal influence between random variables
quantitative measures have been introduced from information theory, such as
time-delayed mutual information and transfer (conditional) entropy. As they
make no assumption about the functional relationship between variables the
latter improve on the more-commonly used measures of time-delayed correlation
\cite{Mwaffo2017} and Granger causality \cite{Bressler2011}, which can capture
only linear functional relationships.

Consider two stationary stochastic processes $X=(\ldots,x_{t-1},x_t,x_{t+1},\ldots)$ and
$Y=(\ldots,y_{t-1},y_t,y_{t+1},\ldots)$ with probability mass functions
$p(x_t)=Pr\{X=x_t\}$ and $p(y_t)=Pr\{Y=y_t\}$, respectively. Their time-delayed
mutual information (TDMI) is given by \cite{Jeong2001}:
\begin{align}
\label{eq: TDMI}
\mathcal{M}_{X \rightarrow Y}(\tau) & = I(X_{t};Y_{t+\tau}) \\
  & = H(Y_{t+\tau}) - H(Y_{t+\tau} | X_{t})
  \nonumber \\
  & = H(X_{t}) - H(X_{t} | Y_{t+\tau})
  \nonumber
  ~,
\end{align}
where $H(Y_{t+\tau})$ and $H(Y_{t+\tau}|X_{t})$ are the Shannon entropy and
conditional entropy, respectively. They measure, in turn, the uncertainty in
$Y_{t+\tau}$ and the uncertainty in $Y_{t+\tau}$ remaining given $X_{t}$
with a delay time $\tau$, respectively. In other words, being their difference
the mutual information $I(X_{t};Y_{t+\tau})$ monitors the reduction
in uncertainty in $Y$'s future knowing $X$'s at a time $t$. Since mutual
information is symmetric, this is also the reduction of uncertainty in $X$'s
present knowing $Y$ at the future time $t+\tau$. The symmetry, though, means
that it cannot be used to infer causal influence since, by assumption, the
future cannot influence present.

In addition, TDMI has a another, perhaps more subtle drawback---when predicting
influence it can be nonzero when two variables have shared history
\cite{Schreiber2000}. That is, the condition $I(X_{t};Y_{t+\tau})>0$ may hold
when variable $Y$ is not directly influenced by variable $X$ but when either
$X$ or $Y$ dynamics contains memory of their past configurations.

Transfer entropy (TE) was introduced to overcome these shortcomings
~\cite{Schreiber2000}: if $X$ influences $Y$, then predicting $Y$'s future
becomes easier after knowing the present of both $X$ and $Y$, compared to only
knowing $Y$'s present. The TE from $X$ to $Y$ takes the form of a conditional
mutual information:
\begin{align}
\label{eqn: TE}
\mathcal{T}_{X\to Y}(\tau) &= I(X_{t};Y_{t+\tau}|Y_{t}) \\
  & =	H(Y_{t+\tau}|Y_{t})-H(Y_{t+\tau}|Y_{t},X_{t} )
  \nonumber
  ~.
\end{align}
That is, $\mathcal{T}_{X\to Y}(\tau)$ is time-delayed mutual information
between $Y$ at time $t+\tau$ and $X$ at time $t$ conditioned by $Y$ at time
$t$. It is the same as subtracting the uncertainty remaining in $Y$ at time
$t+\tau$ given both $X$ and $Y$ at the present time $t$ from that in
$Y_{t+\tau}$ given $Y_{t}$. The latter corresponds to the uncertainty of
$Y_{t+\tau}$ reduced by knowing $X_{t}$ in addition to the knowledge of $Y_{t}$.

$\mathcal{T}_{X\to Y}$ has become one of the standard methods for measuring
statistical influence in classifying leaders and followers
\cite{Butail16,Mwaffo2017,Mwaffo2018,Lord2016,Orange2015,Porfiri2018,Mwaffo2018_2}.
More broadly, since it improves upon TDMI for quantifying asymmetric
relationships, it has become a standard for inferring causal relationships in
many areas of science, including
neuroscience~\cite{Ramirez2021,Vicente2011,Spinney2017,Wollstadt2014,Wibral2011,Wibral2014},
chemistry~\cite{Bauer2006}, human behavior~\cite{Nakayama2017,Porfiri2019}, and
Earth systems~\cite{Delgado2020,Gerken2019,Campuzano2018}.

We note, however, that, like correlation, information-theoretic quantities such
as TDMI and TE are not sufficient in themselves to identify causality. The
latter also requires accounting for the influence of latent or hidden
variables. Since conditioning on the past in computing TE and TDMI are both
finite in time length, each variable's history may act like a hidden variable
that influences outcomes. This, in turn, can lead spurious effects when
estimating information flow, as we will elucidate shortly.

Recently, it was demonstrated for a simple binary system that
$\mathcal{T}_{X\to Y}(\tau) > 0$ can occur even though knowledge of $X_{t}$
alone cannot reduce the uncertainty in $Y_{t+\tau}$~\cite{James2018}.
It was pointed out that, in addition to information intrinsic to reducing
uncertainty in $Y_{t+\tau}$ that comes from knowing $X_{t}$'s present is
independent of $Y_{t}$'s present, TE from $X$ to $Y$ includes information that
reduces the uncertainty in $Y_{t+\tau}$ that comes from knowing $X_{t}$'s
present and $Y_{t}$ simultaneously~\cite{James2016}.

Here, intrinsic information flow is the additional reduction in uncertainty in
$Y$'s future gained from knowing $X$'s present compared to the reduction from
knowing $Y$'s present alone or knowing the present of $X$ and $Y$
simultaneously. Intrinsic mutual information was
proposed~\cite{Maurer1999,James2016} as a measure that avoids including
influence that comes from both the present of $X$ and $Y$ when predicting $Y$'s
future. Note that, while IMI is a specific quantity
and not synonymous with intrinsic information flow, it can be seen as an
attempt to compute intrinsic information flow between two variables.

\subsection{Diagnosing causal influence via intrinsic, shared, and synergistic
informations}
\label{sec: inference_int}

IMI is best appreciated in a cryptographic
setting~\cite{James2016}: Intrinsic information flow between $X$ and $Y$ is
synonymous with information communicated via secret key
agreement~\cite{Maurer1993, Maurer1999} between $X_t$ and $Y_{t+\tau}$ while
$Y_{t}$ is an outside observer. Based on this ansatz, the amount of information
flowing intrinsically from $X$ to $Y$ is equal to the secret key agreement rate
$S(X_{t};Y_{t+\tau}|Y_{t})$. (Defined in Sec.~\ref{Appendix: intrinsic}.) Since
$S(X_{t};Y_{t+\tau}|Y_{t})$ is not computable in practice, intrinsic mutual
information $\mathcal{I}_{X\rightarrow Y}$ is used as a (workable) upper bound
on $S(X_{t};Y_{t+\tau}|Y_{t})$ to monitor information that flows intrinsically
from $X$ to $Y$.

The amount of IMI communicated from process $X$ to
another $Y$ is the infimum of $I(X_{t};Y_{t+\tau}|\overline{Y}_{t})$ taken over
all possible conditional distributions
$p(\overline{Y}_{t}|Y_{t})$ \cite{James2018}:
\begin{align}
\mathcal{I}_{X\rightarrow Y}(\tau):=\mathrm{inf}\Big\{I(X_{t};Y_{t+\tau}|\overline{Y}_{t}): \sum_{y\in Y_{t}}p(X_t,Y_{t+\tau},Y_{t}=y)p(\overline{Y_{t}}|Y_{t}=y)\Big\}
  ~.
\label{eq: intrinsic}
\end{align}
In this, $\overline{Y_{t}}$ is an auxiliary variable used to realize the upper
bound of $S(X_{t};Y_{t+\tau}|Y_{t})$. It satisfies the Markov
property $X_{t}Y_{t+\tau}\rightarrow Y_{t}\rightarrow \overline{Y_{t}}$. Here, $A\rightarrow B$ ($AC\rightarrow B$) signifies that $B$ depends only on $A$ ($A$ and/or $C$) and the infimum is taken over all possible conditional distributions $p(\overline{Y_{t}}|Y_{t})$.

IMI $\mathcal{I}_{X\rightarrow Y}$ represents
uncertainty reduction in $Y$'s future that comes from knowing only $X$'s
present as much as possible under the assumption of the Markov property with
respect to $Y$ and $\overline{Y}$. It should be noted here that this Markov
property has no physical relevance to the actual system $Y$, and is merely an
assumption that restricts the space of the minimization in Eq.~(\ref{eq:
intrinsic}) and so helps to provide an upper bound of
$S(X_{t};Y_{t+\tau}|Y_{t})$. IMI has been found to be
both a convenient and accurate bound on the secret key agreement rate
$S(X_{t};Y_{t+\tau}|Y_{t})$~\cite{Maurer1999,James2016}.

The following relations delineate the importance of $\mathcal{I}_{X\rightarrow
Y}(\tau)$ and its relationship to $S(X_{t};Y_{t+\tau}|Y_{t})$,
$\mathcal{T}_{X\to Y}(\tau)$, and $M_{X\to Y}(\tau)$:
\begin{align}
\label{ineq: I_S}
0 \leq S(X_{t};Y_{t+\tau}|Y_{t}) & \leq  \mathcal{I}_{X\rightarrow Y}(\tau)
  ~, \\
\label{ineq: I_T}
 \mathcal{I}_{X\rightarrow Y} & \leq \mathcal{T}_{X\to Y}(\tau)
  ~,~\text{and} \\
\label{ineq: I_M}
 \mathcal{I}_{X\rightarrow Y} & \leq \mathcal{M}_{X\to Y}(\tau)
  ~.
\end{align}
In effect, Eqs.~(\ref{ineq: I_S})-(\ref{ineq: I_M}) demonstrate that intrinsic
mutual information can be used to compute bounds on the deviations of
$\mathcal{T}_{X\to Y}(\tau)$ and $\mathcal{T}_{X\to Y}(\tau)$ from
$S(X_{t};Y_{t+\tau}|Y_{t})$. That is, whenever equality does not
hold in Eq.~(\ref{ineq: I_T})~(Eq.~(\ref{ineq: I_M})), then there must be a
portion of $\mathcal{T}_{X\to Y}(\tau)$ ($\mathcal{M}_{X\to Y}(\tau)$) that is
not intrinsically coming from $X$. From here on, we set $\tau=1$ and omit
$\tau$ from equations, as it has been shown that $\tau=1$ best captures the
information flow between two particles in the Vicsek model~\cite{Basak2020}.

Once $\mathcal{I}_{X \rightarrow Y}$ has been determined, the shared
information $\sigma_{X\rightarrow Y}$ and synergistic
$\mathcal{S}_{X\rightarrow Y}$ information follow immediately by subtracting it
from TDMI ($\mathcal{M}_{X \rightarrow Y}$) and TE ($\mathcal{T}_{X\rightarrow
Y}$):
\begin{align}
\label{eq: shared}
\sigma_{X\rightarrow Y}
  & = \mathcal{M}_{X \rightarrow Y} - \mathcal{I}_{X\rightarrow Y}
  ~\text{and} \\
\label{eq: synergistic}
\mathcal{S}_{X\rightarrow Y}
  &=  \mathcal{T}_{X\rightarrow Y} -\mathcal{I}_{X\rightarrow Y}
  ~.
\end{align}
From Eqs.~(\ref{ineq: I_S})-(\ref{ineq: I_M}), one sees that:
\begin{align*}
0 & \leq \sigma_{X\rightarrow Y} \leq  \mathcal{M}_{X\rightarrow Y}
  ~\text{and} \\
  0 & \leq \mathcal{S}_{X\rightarrow Y} \leq  \mathcal{T}_{X\rightarrow Y}
  ~.
\end{align*}

Note that $\mathcal{I}_{X\rightarrow Y}$ monitors information coming from
(mostly) $X$ alone to $Y$, since IMI provides an upper
bound on $S(X_{t};Y_{t+\tau}|Y_{t})$. Due to this, $\mathcal{S}_{X\rightarrow
Y} > 0$ ($\sigma_{X\rightarrow Y} > 0$) implies that $\mathcal{T}_{X\rightarrow
Y}$ ($\mathcal{M}_{X\rightarrow Y}$) contains information that comes from $Y$'s
present and that $\mathcal{T}_{X\rightarrow Y}$ ($\mathcal{M}_{X\rightarrow
Y}$) should not be interpreted as information flowing only from $X$ to $Y$ in
those cases.

In the following, based on $\mathcal{I}_{X\rightarrow Y}$ being (mostly)
information coming from only $X$ to $Y$, $\sigma_{X\rightarrow Y}$ is that part
of $\mathcal{M}_{X \rightarrow Y}$ which comes from knowing both variables and
on postulating that it is the information redundant in both $X$ and $Y$, we
referred to it as ``shared' information. Similarly, $\mathcal{S}_{X\rightarrow
Y}$ is \emph{synergistic information}~\cite{James2018} since it is that part of
$\mathcal{T}_{X \rightarrow Y}$ which comes from knowing both variables and
since it arises from simultaneously knowing $X$ and $Y$. Together,
$\mathcal{I}$, $\sigma$, and $\mathcal{S}$ reveal a much more detailed
decomposition of the relationship between $X$ and $Y$ than can be inferred from
$\mathcal{T}$ or $\mathcal{M}$ alone.

To ground this information-theoretic setting, the following shows, using a
modified Vicsek model of collective behavior, that $\mathcal{T}$ and $\mathcal{M}$ can result in misleading interpretation concerning the underlying
actual relationship among individuals. We go on to propose that $\mathcal{I}$,
$\sigma$, and $\mathcal{S}$ provide a firmer interpretation of the relationship
without requiring additional experiments.

\section{Results}

With this background, we now turn to diagnose interactions between individuals
in a collective system and between individuals and the collective.

\subsection{Modified Vicsek Model}

To demonstrate the interpretability of informational modes $\mathcal{I}$, $\sigma$, and
$\mathcal{S}$, we introduce a series of augmented Vicsek models. These extend the original~\cite{Vicsek1995}
with asymmetric interactions and turn on-off the dependence on the present
dynamics of interacting particles.

Consider $N$ particles lying within a square box of length $L$ with periodic
boundary conditions. Particle $i$'s position $\vec{r}_i^{t}$ at time $t$ is
updated over time increment $\Delta t$ according to:
\begin{align}
\vec{r}_i^{t+1}=\vec{r}_i^{t}+\vec{v}_i^{t} \Delta t
  ~,
\label{eqn: position}
\end{align}
where $\vec{v}_i^{t}$ denotes particle $i$'s velocity at time $t$ and
$i=1,2,...,N$. For simplicity, particles have uniform constant speed $v_0$ and
only their orientations $\theta_i$ change.

Particle orientation is updated at each time by taking the weighted average of
the velocity of neighboring particles within a given radius $R$:
\begin{align}
\label{eqn: weight}
\theta_i(t+1) = \langle {\bm \theta}(t)\rangle_{R,\textbf{\textit{w}},\vec{\bm r}^{t}}+\Delta\theta_i
  ~,
\end{align}
where  $\boldsymbol{w}$ is a
nonnegative asymmetric matrix whose $w_{ij}$ element determines the interaction
strength that particle $i$ exerts on particle $j$. $w_{ij}>w_{ji}$ whenever
particle $i$ is a leader and particle $j$ is a follower in our setting.
To model thermal noise $\Delta\theta_{i}$ is a random number uniformly distributed
in the range $[-\eta_0/2, \eta_0/2]$ and is chosen uniquely for each particle
$i$ at each time step. In the original model, the righthand side of~Eq.(\ref
{eqn: weight}) ensured that $\theta_i(t+1)$ resulted from the configurations of
all the particles $j$ (including that of the same particle $i$) within the
circle of radius $R$ centered at $\vec{r}_i^{t}$.

Now, consider modified dynamics that modulate the dependence on $\theta_j(t)$
associated with follower-leader interactions that determine $\theta_i(t+1)$:
the leader influences the follower, but the follower does not influence the
leader; i.e., $w_{\rm LF} > 0$ while $w_{\rm FL}=0$.
\begin{figure}
\includegraphics[width=0.7\linewidth]{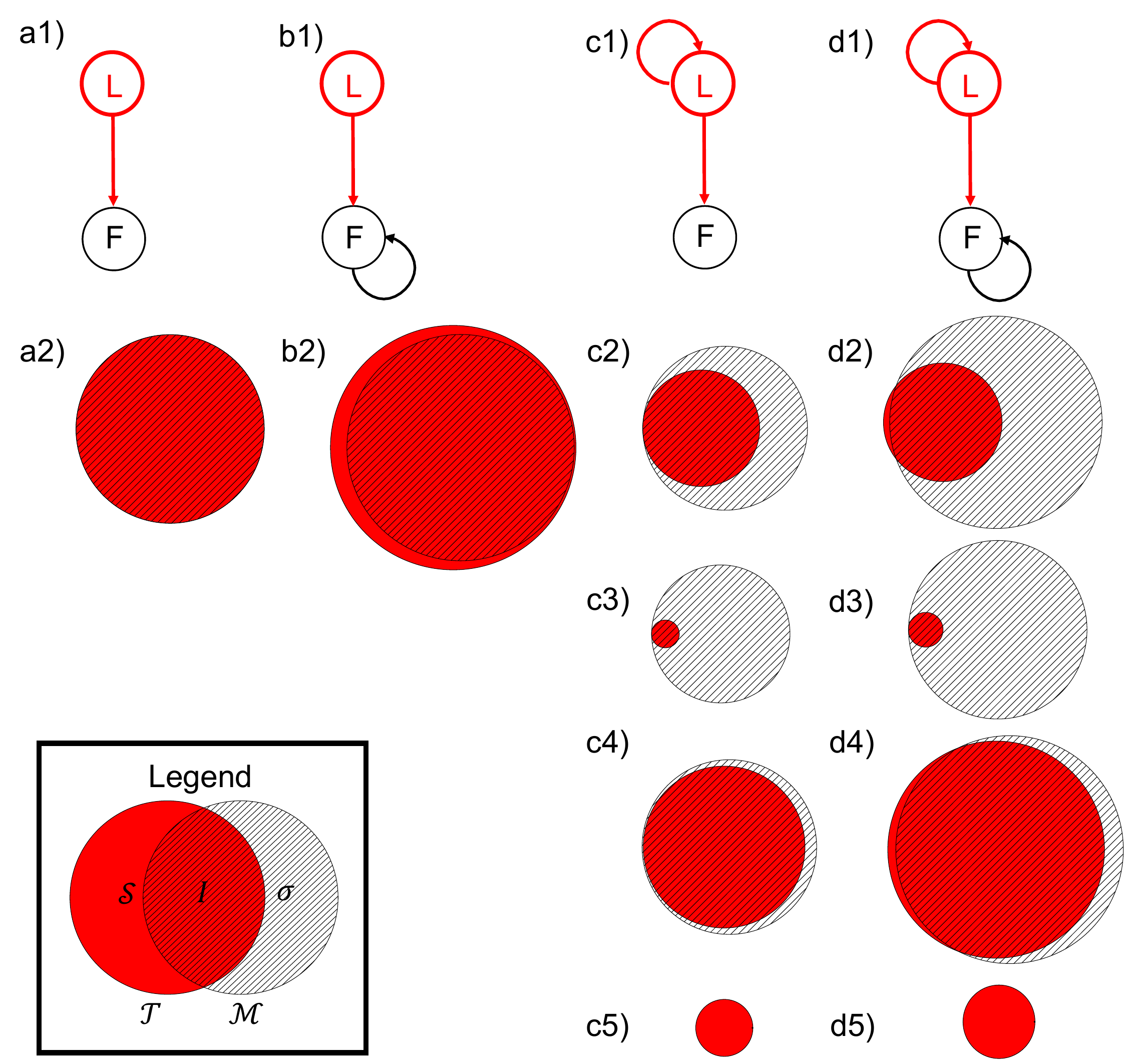}
\caption{Graph representation of interaction types A, B, C, and D and the corresponding Venn diagrams representing information flow integrated over $\eta_0$ ranging from 0 to $2\pi$ and $w_{\rm LF}$ ranging from 1.0 to 10.0.  In all interaction types, the follower depends on the leader but the leader does not depend on the follower, i.e., $w_{\rm LF}>0$ and $w_{\rm FL}=0$. The area of red circle and white striped circle in the Venn diagrams are equal to the (integrated) TDMI $\iint \mathcal{M}(\eta_0,w_{\rm LF})\,d\eta_0\,dw_{\rm LF}$ and the TE $\iint \mathcal{T}(\eta_0,w_{\rm LF})\,d\eta_0\,dw_{\rm LF}$, respectively. The centers of the two circles are determined as follows: first each of the centers is connected with a horizontal line without being overlapped (the TE (TDMI) is located at the left (right)), and then a binary search algorithm was used to find the placement of those circles whose overlapping area is equal to the intrinsic mutual information $\iint \mathcal{I}(\eta_0,w_{\rm LF})\,d\eta_0\,dw_{\rm LF}$ by decreasing the distance between the centers. The part of the red (white striped) circle not overlapped with the white striped (red) circle has area equal to the synergistic information $\iint \mathcal{S}(\eta_0,w_{\rm LF})\,d\eta_0\,dw_{\rm LF}$ (the shared information $\iint \sigma(\eta_0,w_{\rm LF})\,d\eta_0\,dw_{\rm LF}$) (See Legend.) {\bf a1)} Type A. Dynamics of L and F do not depend on their present states. {\bf b1)} Type B Only F depends on its present state. {\bf c1)}  Type C. Only L depends on its present state. {\bf d1)} Type D. Dynamics of L and F both depend on their own present state. In cases where the dynamics of L (F) depends on its present, $w_{\rm LL}$ ($w_{\rm FF}$) = 1. Note that the value of $w_{\rm LL}$ ($w_{\rm FF}$) is inconsequential when the dynamics of leader (follower) do not depend on their present state and $\theta_i(t)$ depends solely on a random number in the interval $[0, 2\pi]$.
%In the Venn diagrams, a white striped circle represents TDMI, a red solid circle represents TE, and their overlapped region represents intrinsic mutual information, while the non-overlapping parts of TDMI and TE represent shared and synergistic information, respectively (see Eqs. 7-8).
{\bf a2)-d2)} Venn diagrams from leader to follower for interaction types A-D.   The information flows from follower to leader for types A and B are negligible, and therefore are not shown. {\bf c3)-d3)} Those from follower to leader for interaction types C-D.  {\bf c4)-d4)}  Those from leader to follower for interaction types C\textprime~-D\textprime~where the leader only remembers its present once every two time steps (see Text). {\bf c5)-d5)} Those from follower to leader for interaction types C\textprime~-D\textprime~. }
\label{fig: venn}
\end{figure}
Figures~\ref{fig: venn}(a1)-(d1) depict possible particle influences in a simple,
two-particle system that determine $\theta_i(t+1)$ in four different cases,
where L and F denote leader and follower, respectively. There, if $A$ or $B$ are
either L or F, then $A \rightarrow B$ signifies that $A$'s present state
influences the $B$'s future state. We vary $w_{\rm LF} \in [1,10]$. We
set $w_{\rm LL} = 1$ and $w_{\rm FF} = 1$ for the models in which the present
state of L (F) influences the future state of L (F). For models in which the
present does not influence the future for the same particle (F or
L)---see Figs.~\ref{fig: venn}(a1) (i.e., L's and F's dynamics),~\ref{fig: venn}(b1) (L's) and \ref{fig: venn}(c1) (F's)---we replace
$\theta_i(t)$ that appears in computing $\langle {\bm \theta}(t)\rangle_{R,\textbf{\textit{w}},\vec{\bm r}^{t}}$ (see $\sum'$ term in Eq.~\ref{S10}) by a random number in the interval $[0,2\pi]$ to
erase any influence from $\theta_i(t)$'s present. For type C and D, in order to address the effect of the history of L,
we also introduced interaction type C\textprime, and D\textprime~which are the same as interactions type C and D respectively, except the future state of L depends on its present only
when time step $t$ is even, and L ``forgets'' its present in its future dynamics as in types A and B whenever $t$ is odd. In types C\textprime~and D\textprime~, the dependence of the future of F on its present are not changed, that is, the future of F does not depend on its present in type C\textprime~and the future of F always depends on its present in type D\textprime~regardless of the value of $t$.

\subsection{Informational modes between leaders and followers }
\subsubsection{Misinterpreting causal influence}

Let us first examine the amounts of TDMI ($\mathcal{M}$) and TE ($\mathcal{T}$)
 shown in Fig.~\ref{fig: venn} for different interaction types, integrated over ranges of both
$w_{\rm LF}$ and $\eta_0$ (the landscapes of $\mathcal{M}$ and $\mathcal{T}$ as a function of $w_{\rm LF}$ and $\eta_0$ are shown in supplemental figures S1-S2). As per the legend in Fig.~\ref{fig: venn}, the white circle with
diagonal shading represents $\mathcal{M}$ and the red circle represents $\mathcal{T}$.
The overlapping region between  $\mathcal{M}$ and  $\mathcal{T}$, the part of
$\mathcal{M}$ which is not overlapping with $\mathcal{T}$, and
the part of $\mathcal{T}$ which is not overlapping with  $\mathcal{M}$ represent $\mathcal{I}$, $\sigma$,
and $\mathcal{S}$, respectively, and are discussed in Sect.~\ref{sec: modes}.

As expected, $\mathcal{M}_{\rm L \rightarrow F}$ and  $\mathcal{T}_{\rm L \rightarrow F}$
in Figs.~\ref{fig: venn}(a2)-(d2),(c4)-(d4), exhibit significant, nonzero values in
all interaction types, since L is influencing F in all cases.
Naively, one expects $\mathcal{M}_{\rm F \rightarrow L}$ and
$\mathcal{T}_{\rm F \rightarrow L}$ to be zero for all cases since F
does not influence L at all. However, there are spurious values of both
$\mathcal{M}_{\rm F \rightarrow L}$ and $\mathcal{T}_{\rm F \rightarrow L}$
in Figs.~\ref{fig: venn}(c3)-(d3), as well as spurious values of  $\mathcal{T}_{\rm F \rightarrow L}$ alone in
Figs.~\ref{fig: venn}(c5)-(d5). Spurious amounts of $\mathcal{T}_{\rm X \rightarrow Y}$ and $\mathcal{M}_{\rm X \rightarrow Y}$
even in cases where X does not influence Y are ignored by a large and growing body of research in quantifying causal relationships.
Section~\ref{sec: modes} elaborates on how decomposing $\mathcal{M}$ and  $\mathcal{T}$ into $\mathcal{I}$, $\sigma$, and $\mathcal{S}$
can improve the interpretation of
%of
information flow using different interaction types as examples.

\subsubsection{Modes of information flow}
\label{sec: modes}
We now interpret by Venn diagrams the different modes of information flow  integrated over ranges of
$w_{\rm LF}$ and $\eta_0$
for each interaction type. The landscapes of $\mathcal{I}$, $\sigma$, and $\mathcal{S}$ as a function of $w_{\rm LF}$ and $\eta_0$ are shown in supplemental figures S3-S5 for types A, B, C, and D, and in supplemental Figures S6 and S7, respectively, for types C\textprime~and D\textprime.
% whose graph representations are shown in Figs.~\ref{fig: venn}(a1),~\ref{fig: venn}(b1),~\ref{fig: venn}(c1)
%and~\ref{fig: venn}(d1).
Since each interaction type corresponds to a special
case of information flow, let us briefly examine each one.

In type A, where neither L nor F depends on its present
(Fig.~\ref{fig: venn}(a1)), only $\mathcal{I}_{\rm L\rightarrow F}$ and no other
type of information flow is observed, as seen in Fig.~\ref{fig:
venn}(a2). In this case, since $\mathcal{M}$ and $\mathcal{T}$ overlap completely, there is no
non-overlapping region of either  $\mathcal{M}$ or $\mathcal{T}$, signifying that $\mathcal{I}=\mathcal{M}=\mathcal{T}$
and $\sigma=\mathcal{S}=0$. Furthermore, all F to L information flow
quantities are equal to zero, thus the Venn diagram for type A from F to L is not shown. Although this may not be a realistic case in systems of interacting agents, since
agents are likely to depend not only on each other but also their history, it is only case demonstrated where either $\mathcal{T}$ or $\mathcal{M}$ can accurately
convey causal relationships between agents.

Figure~\ref{fig: venn}(b1) represents type B where only F depends on
its present in the $\theta_i(t+1)$ dynamics. L to F information flows in this case (Fig.~\ref{fig: venn}(b2)) all increase compared to
the case where F\textquotesingle s dynamics do not depend on its present
(Fig.~\ref{fig: venn}(a2)). This further emphasizes our point that dependence on the present state plays a key role in the
calculation of information flow even when the interactions between individuals
are not intrinsically changing. Notably, the red sliver of non-overlapping region between  $\mathcal{M}$ and $\mathcal{T}$ in  Fig.~\ref{fig: venn}(b2) shows the appearance of
synergistic information $\mathcal{S}_{\rm L\rightarrow F}$, which denotes that part of $\mathcal{T}_{\rm L\rightarrow F}$ is not intrinsically coming from L. Since F depends on its past, the simultaneous knowledge of the present state of F and L provide more predictive power than knowing either the present state of F or L alone.
F to L information flows in this case are again equal to zero, since the leader has no history to share with the follower.

Figure~\ref{fig: venn}(c1) represents type C where only L depends on its present.
In this case, a significant amount of $\sigma_{\rm L\rightarrow F}$ appears
due to the dependence of the present state of both L and F on the past state
of L, as shown in Fig.~\ref{fig: venn}(c2). L imparts information from its history
on to the future dynamics of F, and meanwhile, this information is already contained
in the present state of F due to the present state of F\textquotesingle s
dependence on that same history.  In contrast to types A and B, there is a significant amount of
shared history between F and L, as shown in Fig.~\ref{fig: venn}(c2). As in the binary
system proposed by Schreiber~\cite{Schreiber2000} having
the same graph representation as~Fig.\ref{fig: venn}(c1), there
exists a significant amount of $\mathcal{M}_{\rm F \rightarrow L}$ even though there is no direct interaction
in that direction. By decomposing $\mathcal{M}_{\rm F \rightarrow L}$ into  $\mathcal{I}_{\rm F \rightarrow L}$
and $\sigma_{\rm L\rightarrow F}$, we quantitatively show that the spurious amount of  $\mathcal{M}_{\rm F \rightarrow L}$  is coming solely from
$\sigma_{\rm L\rightarrow F}$ and is thus not intrinsically coming from F. Transfer entropy
 $\mathcal{T}$ was introduced to reconcile this issue \cite{Schreiber2000}. $\mathcal{T}_{\rm F \rightarrow L}$
does in fact reduce the amount information flow in that direction in our model, given that
$\mathcal{T}_{\rm F \rightarrow L}$ is significantly less than $\mathcal{M}_{\rm F \rightarrow L}$
in Fig.~\ref{fig: venn}(c3).

Why then, is $\mathcal{T}_{\rm F \rightarrow L}$, and more importantly,
$\mathcal{I}_{\rm F \rightarrow L}$ not equal to zero in type C (Fig.~\ref{fig: venn}(c3))?
Surely, information is not intrinsically flowing from F to L in this case since there
is no direct link from F to L in Fig.~\ref{fig: venn}(c1). The reason
is that the history of L acts as a hidden variable, imparting information onto
both L and F. To verify this,
% claim,
we have introduced interaction type C\textprime,~which is the same
as interaction type C (Fig.~\ref{fig: venn}(c1)) except the future state of L depends on its present only
when time step $t$ is even, and L ``forgets'' its present in its future dynamics when time step $t$ is odd.
Although the present state of L can still act as a hidden variable imparting information on both L and F,
the computation of $\mathcal{I}_{\rm F \rightarrow L}$ conditions on the present state of L and also
minimizes the uncertainty
%reduction
coming from the present state of L as much as possible. Therefore
it is not possible for the history of L to have any effect in type on the value of $\mathcal{I}_{\rm F \rightarrow L}$ in type C\textprime~since
it remembers at most one time step of its
past history, which is being minimized in the computation of  $\mathcal{I}$. The Venn diagram from follower to leader in
this case is shown in Fig.~\ref{fig: venn}(c5). Note that $\mathcal{T}_{\rm F \rightarrow L}$ still exists in
this case, but only in the form of synergistic information $\mathcal{S}_{\rm F \rightarrow L}$ as $\mathcal{I}_{\rm F \rightarrow L}=0$.
Recall that the existence of $\mathcal{S}_{\rm F \rightarrow L}$, which is a part of transfer entropy, implies that simultaneous knowledge of the present states of L  and F allows for an improvement on the prediction of the future of L compared to individual knowledge of the present states of L or F alone. Why does this happen under $\mathcal{I}_{\rm F \rightarrow L}=0$?
Here we explain the origin of the existence of synergistic information intuitively. For interaction type C',  $\theta_{\rm F}(t+1)$ always results from $\theta_{\rm L}(t)$ irrespective of time step, i.e.,  $\theta_{\rm F}(t+1)=\theta_{\rm L}(t)$, in this two body model system. In turn, the present configuration of L, $\theta_{\rm L}(t)$, affects its future, i.e., $\theta_{\rm L}(t+1)=\theta_{\rm L}(t)$ and hence $\theta_{\rm F}(t+1)=\theta_{\rm L}(t+1)$, every {\it even} time step $t$. However, $\theta_{\rm L}(t+1)$ is taken from $[0:2\pi]$ randomly to reset its history every {\it odd} time step $t$,  i.e., $\theta_{\rm F}(t+1) \ne \theta_{\rm L}(t+1)$. That is, simultaneous knowledge of the states of L and F reduces
 uncertainty about whether time step is even or odd, and, therefore, simultaneous knowledge of the present states of L and F improves the prediction power of the future of
L compared to solely individual knowledge of the present state of L and F (See more in detail in Supplemental Information Sect. II).

Finally, Fig.~\ref{fig: venn}(d1) represents type D, which is perhaps the most intuitive case in
typical systems, where both entities depend on their present state in the
$\theta_i(t+1)$ dynamics. $\mathcal{M}$, $\mathcal{T}$, and $\sigma$ are greater than zero for similar reasons as they are in types B and C for both leader to follower and follower to leader, and
$\mathcal{S}_{\rm L \rightarrow F}$ but not $\mathcal{S}_{\rm F \rightarrow L}$ is greater than zero for similar reasons as well. Type D\textprime~is the same as C\textprime, except F depends on its present dynamics as in type D. As in type C\textprime, the only type of information flow from F to L in type D\textprime~is $\mathcal{S}_{\rm F \rightarrow L}$, and for similar reasons. Thus type D, representing the most typical types of multi-agent systems, contains a rich profile of information flows which we have explained by analyzing types A, B, C, C\textprime  and D\textprime.

\subsection{Leader and Group of Followers}

The analysis up to
%this point
now addressed only pairwise interactions. This is in
accord with the theoretical basis of the information measures used; for
example, $\mathcal{T}_{X\rightarrow Y}(\tau)$ in Eq.~(\ref{eqn: TE}). The
measures generalize straightforwardly to account for additional time
series, say, of a third particle (or agent); see, for example, the causation
entropy~\cite{Lord2016}. Suppose the third variable $Z$, in addition to $X$ and $Y$, are each symbolized by $m$ discrete values. 
Then, for example, the dimension of the probability distribution $p(Y_{t+1},X_t,Z_t)$ is $m^3-1$ ($-1$ is because of probability normalization). 
This means that, the more the number of additional variables to be conditioned on 
%additional variables
increases, the more the dimension of the probability distribution required for computing
the measures grows exponentially with respect to the number of additional variables. This requires increasingly large amounts of data to properly
sample. Therefore, in multi-agent systems it is not usually feasible to
condition on all or even a few other agents that interact with a given
agent. In addition, even if additional variable(s) that indirectly affect(s) interactions between $X$ and $Y$ exist, it is nontrivial to look for such indirect `cause.' 
Such hidden variables may be another agent entity, some past memory of the process of $X$ and/or $Y$ longer than being taken into account in the elucidation of TE, or something else. 

% find conditioning on additional variables introduces extra dimensions
% to the number of quantities that needed to be computed, for example $n^3$ ($ {}_nC_3 $)
% when conditioning on a third variable and there are $n$ agents.}

Nonetheless, estimating two-agent information measures has proven useful for
monitoring influence in systems having more than two
agents~\cite{Butail2016,Mwaffo2017,Mwaffo2018,Orange2015,Jeong2001}. We will
now show how measuring $\mathcal{I}$, $\sigma$, and $\mathcal{S}$ gives marked
improvements even in these admittedly-approximate settings.

\begin{figure}
\includegraphics[width=0.4\linewidth]{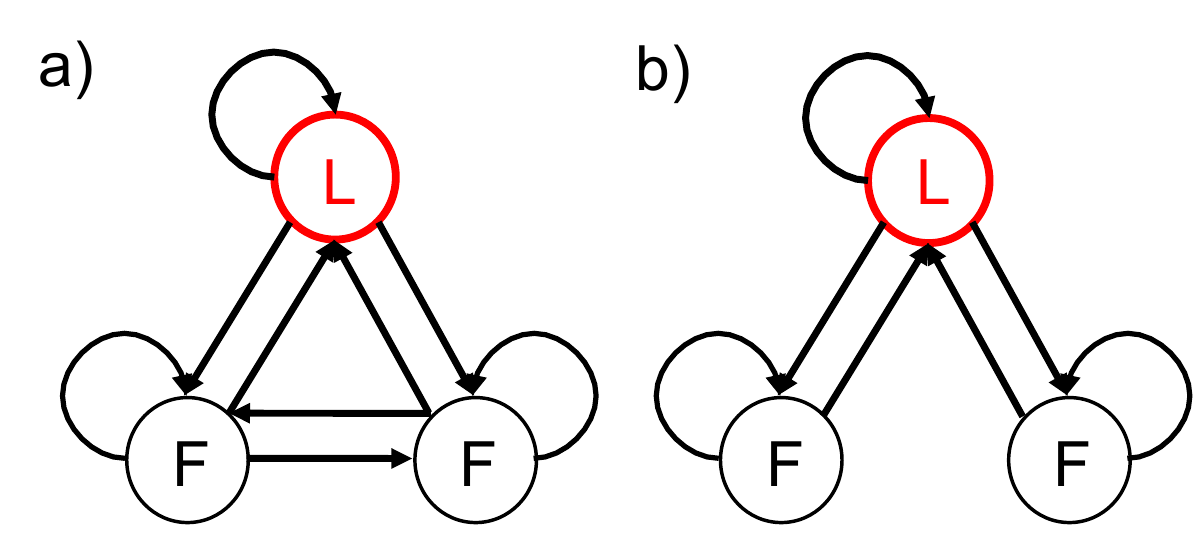}
\caption{Three-agent interaction diagrams: a) In Model A a leader influences
	both followers, both followers influence L, and followers
	influence each other. b) Model B is similar, but followers cannot influence
	each other. Weights are asymmetric: $w_{\rm LF}$ (leader to followers)
	is greater than $w_{ \rm FL}$ and $w_{\rm FF}$. We set $w_{\rm LF}$=4
	and $w_{\rm FL}=w_{\rm FF}=1$.
	}
\label{fig: many}
\end{figure}

Now, consider a collective in which L and F mutually interact
with one another, but under model A followers also directly interact with each
other and under model B they do not. See, for example, Fig.~\ref{fig: many} for the case of three agents. In the following discussion,
there is one leader agent and the number of follower agents $N_{\rm F}$ is varied. L refers to the leader and F refers to a particular follower.

\begin{figure}
\includegraphics[width=\linewidth]{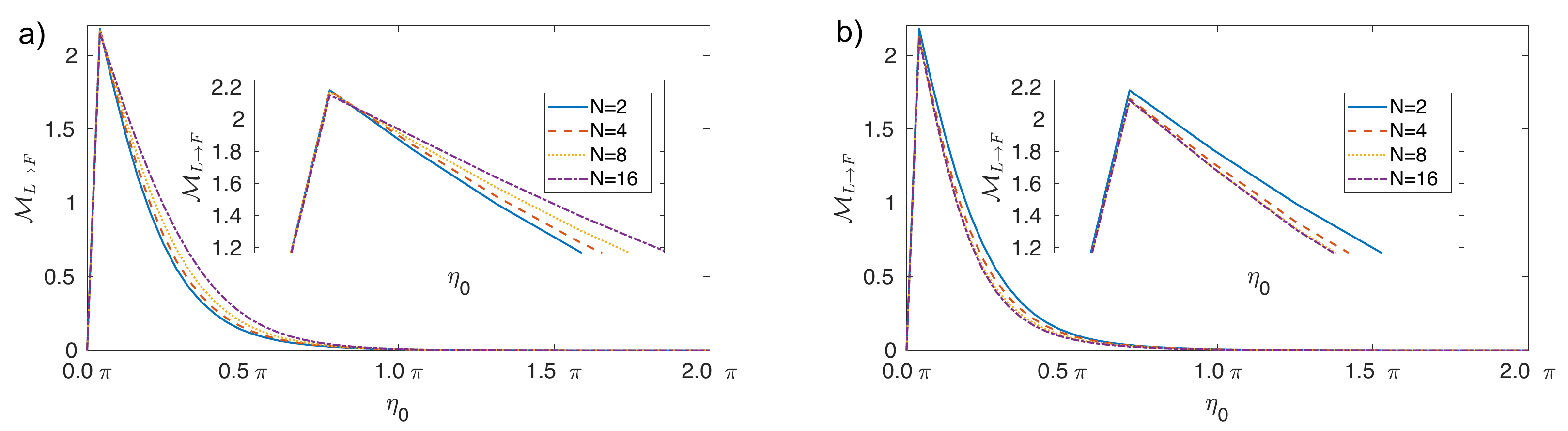}
\caption{$\mathcal{M}$ as a function of noise level $\eta_0$ with$N_{\rm F}=$ 1,
	(blue), 3 (red), 7 (orange) and 15 (purple) where the
	number of leaders is always one. a)  $\mathcal{M}_{{\rm L} \rightarrow {\rm
	F}}$ for model A. b) $\mathcal{M}_{{\rm L} \rightarrow {\rm F}}$ for
	model B. 	}
\label{fig:TDMImany}
\end{figure}

\begin{figure}
\includegraphics[width=\linewidth]{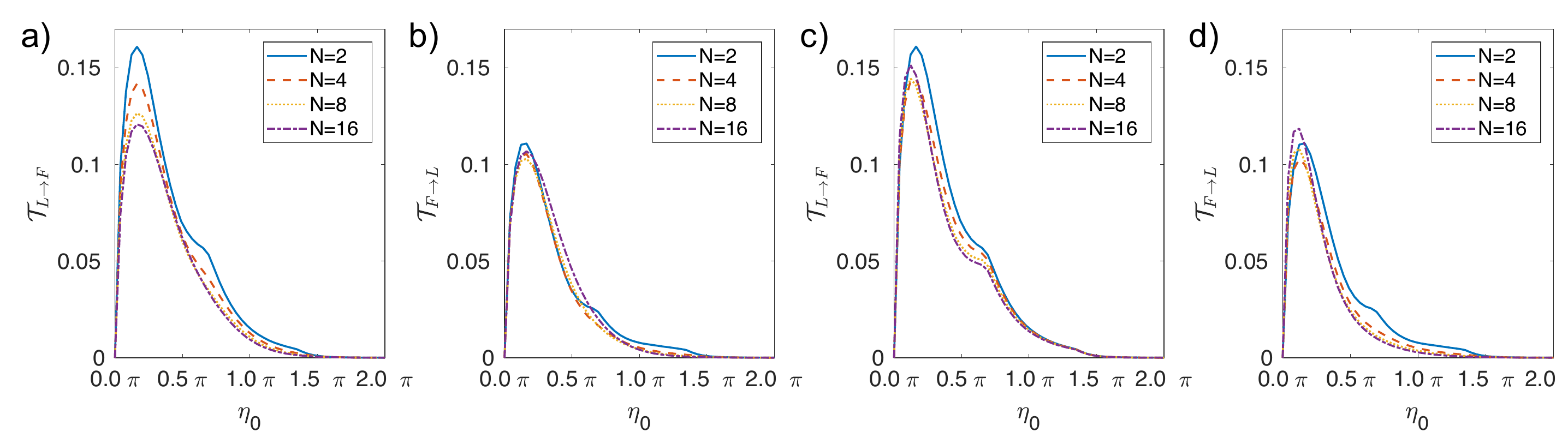}
\caption{$\mathcal{T}$ as a function of noise level $\eta_0$ with $N_{\rm F}=$ 1,
	(blue), 3 (red), 7 (orange) and 15  (purple) where the
	number of leaders is always one. a)  $\mathcal{T}_{{\rm L} \rightarrow {\rm
	F}}$ for model A. b) $\mathcal{T}_{{\rm F} \rightarrow {\rm L}}$ for
	model A. c)  $\mathcal{T}_{{\rm L} \rightarrow {\rm
	F}}$ for model B. d) $\mathcal{T}_{{\rm F} \rightarrow {\rm L}}$ for
	model B.  	}
\label{fig:TEmany}
\end{figure}

\begin{figure}
\includegraphics[width=\linewidth]{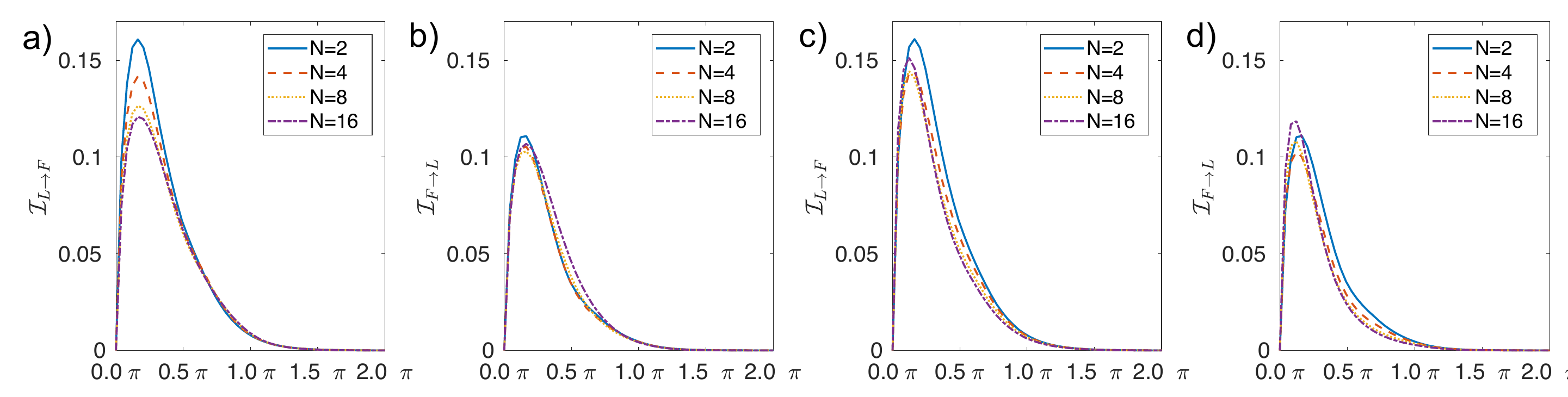}
\caption{$\mathcal{I}$ as a function of noise level $\eta_0$ with $N_{\rm F}=$ 1,
	(blue), 3 (red), 7 (orange) and 15  (purple) where the
	number of leaders is always one. a)  $\mathcal{I}_{{\rm L} \rightarrow {\rm
	F}}$ for model A. b) $\mathcal{I}_{{\rm F} \rightarrow {\rm L}}$ for
	model A. c)  $\mathcal{I}_{{\rm L} \rightarrow {\rm
	F}}$ for model B. d) $\mathcal{I}_{{\rm F} \rightarrow {\rm L}}$ for
	model B.  		}
\label{fig:INTmany}
\end{figure}

\begin{figure}
\includegraphics[width=\linewidth]{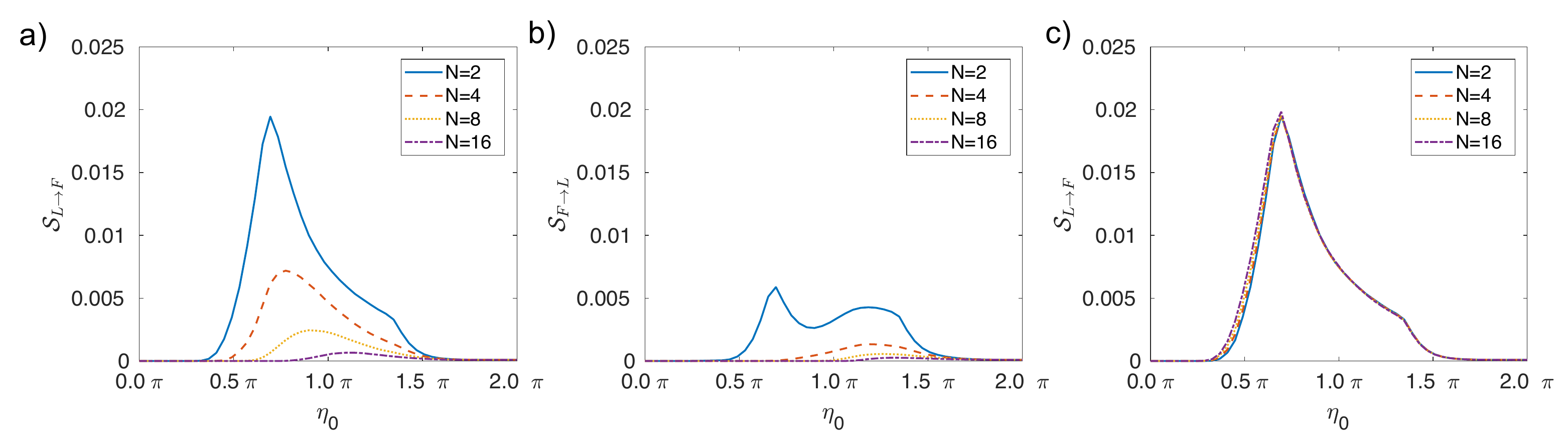}
\caption{$\mathcal{S}$ as a function of noise level $\eta_0$ with $N_{\rm F}=$ 1,
	(blue), 3 (red), 7 (orange) and 15 (purple) where the
	number of leaders is always one. a)  $\mathcal{S}_{{\rm L} \rightarrow {\rm
	F}}$ for model A. b) $\mathcal{S}_{{\rm F} \rightarrow {\rm L}}$ for
	model A. c)  $\mathcal{S}_{{\rm L} \rightarrow {\rm
	F}}$ for model B. 	}
\label{fig:SYNmany}
\end{figure}

Figures~\ref{fig:TDMImany}(a) and~\ref{fig:TDMImany}(b) display $\mathcal{M}_{{\rm L} \rightarrow {\rm F}}$ for model A and model B, respectively. The plots of $\sigma$ (See supplemental Fig.~S9) are almost indistinguishable from those of $\mathcal{M}$, indicating that a majority of $\mathcal{M}$ is actually coming from $\sigma$, which is due to shared history between L and F. As has been established for the Vicsek model~\cite{Vicsek1995}, cohesive behavior increases as a function of density. Here,
%, and therefore
$\mathcal{M}_{{\rm L} \rightarrow {\rm F}}$ and $\sigma_{{\rm L} \rightarrow {\rm F}}$ increase as a function of $N_{\rm F}$ in model A. Model B, however, is not the same as the original Vicsek model in that followers do not interact with each other, and therefore $\mathcal{M}_{{\rm L} \rightarrow {\rm F}}$ and $\sigma_{{\rm L} \rightarrow {\rm F}}$ decrease as a function of $N_{\rm F}$, since the inclusion of additional agents which are not interacting decreases the overall cohesion between
 the present state of L and the future state of F. The plots of $\mathcal{M}_{{\rm F} \rightarrow {\rm L}}$ for model A and B are not shown as they
 are not distinguishable by eye from those of $\mathcal{M}_{{\rm L} \rightarrow {\rm F}}$ (See supplemental Fig.~S8).

Figures~\ref{fig:TEmany}(a)-(d) show $\mathcal{T}_{{\rm L} \rightarrow {\rm F}}$ as a function of $\eta_0$ for
model A, $\mathcal{T}_{{\rm F} \rightarrow {\rm L}}$ for model A,
$\mathcal{T}_{{\rm L} \rightarrow {\rm F}}$ for model B, and $\mathcal{T}_{{\rm F} \rightarrow {\rm L}}$ for model B, respectively. At
 $\eta_0=0$, agent movements quickly reach a regular parallel flow
independent of initial
%states
coordinates and velocities, and thus any information about
%so therefore
their present orientations are
negligible (on average) in predicting the others' orientational motions (See supplemental video 1.).
 In practice, all agents are subject to finite noise due to their environment (represented here by thermal fluctuation).
Gradual decreases of $\mathcal{T}$ as $\eta_0$ increases
simply arise from this natural stochasticity. In both model A and model B there are small bumps
in $\mathcal{T}_{{\rm L} \rightarrow {\rm F}}$ and
$\mathcal{T}_{{\rm F} \rightarrow {\rm L}}$ at
%around
$\eta_0\simeq0.7\pi$, but
%however
the striking difference is that the bumps clearly decrease as a function of $N_{\rm F}$ from L to F and F to L  in model A (Figs.~\ref{fig:TEmany}(a) and~\ref{fig:TEmany}(b)), and from F to L in model B~(Fig.~\ref{fig:TEmany}(d)), but not from L to F in model B (Fig.~\ref{fig:TEmany}(c)).

The existence of bumps in $\mathcal{T}$ at
%around
$\eta_0\simeq0.7\pi$ and the difference in their behavior between model A and model B can be explained by decomposing  $\mathcal{T}$ into  $\mathcal{I}$ and  $\mathcal{S}$. Figures~\ref{fig:INTmany}(a)-(d) show $\mathcal{I}_{{\rm L} \rightarrow {\rm F}}$ as a function of $\eta_0$ for model A, $\mathcal{I}_{{\rm F} \rightarrow {\rm L}}$ for model A,
$\mathcal{I}_{{\rm L} \rightarrow {\rm F}}$ for model B, and $\mathcal{I}_{{\rm F} \rightarrow {\rm L}}$ for model B, respectively. While the overall trend is very similar to that of $\mathcal{T}$ in Fig.~\ref{fig:TEmany},
%however,
$\mathcal{I}$ does not contain bumps at $\eta_0\simeq 0.7\pi$. Thus the bumps in $\mathcal{T}$ are explained solely by $\mathcal{S}$. Figures~\ref{fig:SYNmany}(a)-(c) show $\mathcal{S}_{{\rm L} \rightarrow {\rm F}}$ as a function of $\eta_0$ for model A, $\mathcal{S}_{{\rm F} \rightarrow {\rm L}}$ for model A, and
$\mathcal{S}_{{\rm L} \rightarrow {\rm F}}$ for model B, respectively ($\mathcal{S}_{{\rm F} \rightarrow {\rm L}}$ is indistinguishable by eye from  $\mathcal{S}_{{\rm F} \rightarrow {\rm L}}$ for model B, and therefore is shown in supplemental Fig.~S10). As an overall trend, $\mathcal{S}$ is negligibly small when $\eta_0<0.4\pi$, which means that since the
%state
configuration of F (L) is not varying by a large amount from time to time, the simultaneous knowledge of L and F does not decrease the uncertainty in F (L) more than knowing the current configuration  of L or F alone. At intermediate values of  $\eta_0$, simultaneous knowledge of L and F become relatively more important, and at high values of  $\eta_0$, the simultaneous knowledge of L and F has no predictive power as the dynamics is dominated by thermal noise.  As $N_{\rm F}$ increases in model A,  $\mathcal{S}_{{\rm L} \rightarrow {\rm F}}$ and  $\mathcal{S}_{{\rm F} \rightarrow {\rm L}}$ both decrease, as the future configuration  of F (L) depends on more other agents and relies less on the simultaneous knowledge of L or F alone.
 Therefore, increasing $N_{\rm F}$ decreases the likelihood that simultaneously knowing the configuration of
F and L has any additional predictive power on L or F. In Model B, however, F is not affected by other followers and therefore  $\mathcal{S}_{{\rm L} \rightarrow {\rm F}}$ remains largely unchanged as a function of $N_{\rm F}$.

 \section{Conclusion}

We investigated a series of model systems based on the Vicsek Model of
collective motion to explore the effect of interaction protocols on the
distinct modes of information flow. In theory, one would condition on all
variables, as well as history, to fully interpret mutual relationships among
agents in a collective. At present this is not practical. Instead, our task was
to acquire detailed and correct interpretations under the constraint of limited
measurements---specifically, pairwise interactions among agents.

We observed that the intrinsic information between $X$ and $Y$ dominates
whenever there is only a link from $X$ to $Y$ and no direct link between $Y$ to
$X$ or from $Y$ to itself. However, a small amount of intrinsic information can
still be observed when there is no direct link from $X$ to $Y$, as in the case
where $X$ is a follower with memory and $Y$ is a leader. We noted that this was
due to the effect of memory. We also found that shared information dominates
when there is memory shared between particles due to their interactions and
memory of their pasts. Synergistic information dominates when present knowledge
of $X$ or $Y$ alone cannot predict the future state of $Y$ by itself, but
knowing both the present of $X$ and $Y$ simultaneously does. One of the most
striking consequences in our analysis of this multi-agent system was that
decomposing transfer entropy into intrinsic information and synergistic
information flows enabled us to correctly interpret the ``bump'' observed in
transfer entropy as a function of noise level. Notably, from that one can
infer how each follower interacts with each other in the collective.

Based on the model systems and their corresponding information flows, one can
deduce which information measure is more appropriate based on the physical
problem being addressed. In leader-follower classification, for example,
transfer entropy is often used. However, when one does not expect to find
significant synergistic or shared flows, it equals time-delayed mutual
information. The latter is then a better choice since it does not require
additional conditioning that increases the dimension of the probability
distribution that must be well-sampled. In cases where synergistic flow is
dominant, one may consider separating intrinsic and synergistic flows instead
of computing just transfer entropy. This results in a much richer feature space
for classification. In general, computing intrinsic, shared, and synergistic
flows should perform better or at least as well as transfer entropy and
time-delayed mutual information in classification. Future work will verify
these claims and elucidate exactly in which scenarios we expect each mode of
information flow to be effective in classifying leaders and followers.

\acknowledgments

We thank Profs. J. Green, M. Toda, H. Teramoto and K. Tabata for valuable
discussions. TK and JPC thank the Telluride Science Research Center for its
hospitality during visits. This work was supported by a Grant-in-Aid for
Scientific Research on Innovative Areas ``Singularity Biology (No.8007)''
(18H05413), MEXT, and by JSPS (No. 25287105 and 25650044 to T.K.), and
JST/CREST (No. JPMJCR1662 to T.K.). It is also based upon work supported by, or
in part by, the U. S. Army Research Laboratory and the U. S. Army Research
Office under grants W911NF-18-1-0028 and W911NF-21-1-0048.
\section{Methods}

\subsection{Defining Information Flow}
\label{Appendix: intrinsic}
In this section we will construct our measure of intrinsic information flow.
We will start with a broader understanding of information flow, and then narrow it until we arrive at our goal.
To begin, information flow from a time series $X$ to a time series $Y$ must exist in both the behavior of $X$ at time $t$ and the later behavior of $Y$ at time $t + \tau$:
\begin{align*}
    \mathcal{I}(X_{t} ; Y_{t + \tau})=\sum_{x_t,y_{t+\tau}}p(x_t,y_{t+\tau})\log_2\frac{p(x_t,y_{t+\tau})}{p(x_t)p(y_{t+\tau})}~,
\end{align*}
a quantity known as the time-delayed mutual information.
%\emph{time-delayed mutual information}.

As pointed out by Schreiber~\cite{Schreiber2000}, there are many reasons why $X_{t}$ and $Y_{t+\tau}$ might share information.
Firstly, both $X$ and $Y$ may be synchronized, and so $X_{t}$ predicts $Y_{t+\tau}$ in the same fashion that $Y_{t}$ would, and so it would be disingenuous to attribute that shared information to information flow.
Similarly, $X$ and $Y$ may be jointly influenced by a third system $Z$, and so there is no direct, or even indirect, information flow from $X$ to $Y$.
Schreiber referred to these two situations as the two time series being correlated via \emph{common history} and \emph{common input signals}, and proposed discounting these influences from the time-delayed mutual information via conditioning:
\begin{align*}
    \mathcal{I}(X_{t} ; Y_{t + \tau} | Y_{t})=\sum_{x_t,y_t,y_{t+\tau}}p(y_{t+\tau},y_{t},x_t)\log_2 \left(\frac{p(y_{t+\tau}|y_t,x_t)}{p(y_{t+\tau}|y_t)}\right)~,
\end{align*}
a quantity known as the transfer entropy.
This does overcome the stated weakness of the time-delayed mutual information, and removes the information shared between $X_t$ and $Y_{t + \tau}$ which is also present in $Y_t$.
The transfer entropy can also be modified to discount the information also in a simultaneous third variable:
\begin{align*}
    \mathcal{I}(X_{t} ; Y_{t + \tau} | Y_{t}, Z_{t})=\sum_{x_t,y_t,y_{t+\tau}}p(y_{t+1},y_{t},x_t,z_t)\log_2 \left(\frac{p(y_{t+1}|y_t,x_t,z_t)}{p(y_{t+1}|y_t,z_t)}\right)~.
\end{align*}

Conditioning on variables, however, is not a purely subtractive operation.
That is, the following relation does not hold:
\begin{align*}
    \mathcal{I}(X ; Y | Z) \leq \mathcal{I}(X ; Y)~.
\end{align*}
Rather, conditioning can \emph{increase} the information shared by two variables.
This phenomena is known as conditional dependence~\cite{husmeier2005introduction}, and is perhaps best exemplified by the following distribution, where $X$, $Y$, and $Z$ are binary random variables and the events in which an even number of them take on the value 1 have equal probability:\\
\begin{table}[h!]
    \centering
    \begin{tabular}[t]{cccc}
        \multicolumn{4}{c}{\textsc{Xor}} \\
        \toprule
        $X$ & $Y$ & $Z$ & $\mathrm{Pr}$ \\
        \midrule
        0   & 0   & 0   & $\frac{1}{4}$ \\
        0   & 1   & 1   & $\frac{1}{4}$ \\
        1   & 0   & 1   & $\frac{1}{4}$ \\
        1   & 1   & 0   & $\frac{1}{4}$ \\
        \bottomrule
    \end{tabular}
\end{table}
In this distribution, any pair of variables are independent:
\begin{align*}
    \mathcal{I}(X ; Y) = \mathcal{I}(X ; Z) = \mathcal{I}(Y ; Z) = 0~,
\end{align*}
yet each of those pairs conditioned on the third variable is highly correlated:
\begin{align*}
    \mathcal{I}(X ; Y | Z) = \mathcal{I}(X ; Z | Y) = \mathcal{I}(Y ; Z | X) = 1~.
\end{align*}
This is because knowing the value of, for example, $X$, does not allow you to infer the value of $Y$, but if conditioned on (again, for example) $Z = 0$, suddenly we know that whatever value $X$ takes, $Y$ must also.
This is conditional dependence in its purest form: $X$ and $Y$ are independent, but given $Z$ they are perfectly correlated.

This brings us back to the transfer entropy.
As it is based on a particular conditional mutual information, conditioning on $Y_{t}$ can induce a correlation between $X_{t}$ and $Y_{t + \tau}$ which does not exist without it.
In order to overcome this weakness in the transfer entropy, we propose taking a step back and considering the problem from a slightly different perspective.
We seek an operational understanding of the information shared by $X_{t}$ and $Y_{t + \tau}$, removing the influences of common history and common input signals, without introducing other forms of correlation.
To do this we appeal to the cryptographic flow ansatz~\cite{James2018}, which states that intrinsic information flow exists when $X_{t}$ and $Y_{t + \tau}$ can agree upon a secret while $Y_{t}$ acts as an eavesdropper, and furthermore that the intrinsic information flow is quantified as the rate of secret sharing existing between the two.
In essence, this means that information shared by $X_{t}$ and $Y_{t + \tau}$ can only be definitively attributed to flow from $X_{t}$ to $Y_{t + \tau}$ if there is no way that information can be reconstructed or derived by $Y_{t}$.

In order to practically apply the cryptographic flow ansatz, we utilize a relatively easily computable upper bound, termed as intrinsic mutual information:
\begin{align}
    \mathcal{I}(X ; Y \downarrow Z) \equiv \inf_{p(\bar{z}|z)} \mathcal{I}(X ; Y | \bar{Z})~.
\label{Eq:intrinsic_def}
\end{align}
Effectively, this bound simply states that the information shared by $X$ and $Y$ which is inaccessible to $Z$ is bound from above by the conditional mutual information between $X$ and $Y$ given all possible variables that can be constructed from $Z$.
As an eavesdropper, $Z$ is not limited to what
%they \tk{(=$X$ and $Y$ or =$Z$? I read as $Z$ but $Z$ is an icon of an evesdropper, so sigle)}
he/she directly observes, but can also transform or process his/her
%their
observations as he/she
%they
so chooses.
Since the eavesdropper does not have direct access to either $X$ or $Y$,
%their\tk{(=($X$ and $Y$)'s or =$Z$'s ?)}
his/her transformed variable $\bar{Z}$ cannot have been directly influenced by those variables $X$ and $Y$, and so are given by the conditional distribution $p(\bar{z}|z)$, a statistical mapping from $z$ to $\bar{z}$.
Equivalently, the Markov chain $XY \rightarrow Z \rightarrow \bar{Z}$ holds  ($A\rightarrow B$ ($AC\rightarrow B$) means that $B$ depends only on $A$ ($A$ and/or $C$)): although the eavesdropper $Z$ has no direct access to information shared by $X$ and $Y$, $Z$ outputs/chooses a certain value $z$ by somehow inferring present communications between $X=x$ and $Y=y$, expressed by $XY \rightarrow Z$, and the auxiliary variable $\bar{Z}$ constructs all possible values $\bar{z}$ from $Z=z$, expressed by $Z \rightarrow \bar{Z}$. This Markov chain ensures that the transformed variable $\bar{Z}$ cannot directly be influenced by $X$ and $Y$.
At two extremes---$\bar{z}$ is a constant (e.g., zero) and $\bar{z}$ is identical to $z$---we recover the mutual information and the conditional mutual information:
\begin{align}
  \forall z, p(\bar{z} = 0|z) = 1 &\implies \mathcal{I}(X ; Y | \bar{Z}) = \mathcal{I}(X ; Y) \\
  \forall z, p(\bar{z} = z|z) = 1 &\implies \mathcal{I}(X ; Y | \bar{Z}) = \mathcal{I}(X ; Y | Z)~,
\end{align}
and we therefore have the following inequalities from Eq. \ref{Eq:intrinsic_def}:
\begin{align}
  \mathcal{I}(X ; Y \downarrow Z) &\leq \mathcal{I}(X ; Y) \\
  \mathcal{I}(X ; Y \downarrow Z)   &\leq \mathcal{I}(X ; Y | Z)~,
\end{align}
which allows us to consider the intrinsic mutual information in fact a subtractive operation, isolating a component of the information shared by $X$ and $Y$, but from which much of the influence of $Z$ has been removed~\cite{renner2003new} without creating conditional dependence.

The calculation of the intrinsic mutual information, while not trivial, is not particularly difficult.
While the optimization over $p(\bar{z}|z)$ is not convex, the optimization space is finite because the cardinality of $\bar{z}$ can be bound by the cardinality of $z$~\cite{christandl2003property}.
The object of optimization is then a $|z| \times |z|$ row stochastic matrix where the $i, j$th entry is $p(\bar{z} = j | z = i)$.
Global optimization techniques, such as basin hopping, can then be utilized to find the global minima.
In basin hopping, an initial condition is proposed, and the local minima found through standard gradient-based techniques; then a step in the optimization space is taken and the local minima found again.
This is repeated some number of times, and the least of the found local minima is presumed to be the global minima.
This is the technique used in the \texttt{dit} information theory package~\cite{James2018_2}, which was used to perform the calculations in this paper.

In short, this measure builds upon the transfer entropy, producing a new metric which comes significantly closer to the transfer entropy's stated goal of removing the effects of common history and common input signals, but without introducing the possibility of conditional dependence.
This is accomplished by appealing to the field of information theoretic cryptography, and drawing parallels between secret key agreement and the scientific issue of attributing information present in $Y_{t + \tau}$ to $X_{t}$, and $X_{t}$ alone. 

\subsection{Computing information flow measures}
The computation of $\mathcal{T}$, $\mathcal{M}$, $\mathcal{S}$, and $\sigma$ were performed
% are done
as follows. First, the orientations $\theta_{\rm L}$ and $\theta_{\rm F}$ are computed as described by Eq.~\ref{eqn: weight} and detailed in Sect.~\ref{sec: vicsek_detail} up to time $T=2\times 10^6$ for 20 sets of initial conditions. The values of $\theta_{\rm L}$ and $\theta_{\rm F}$  are then discretized by binning them into 6 bins on the interval $[0, 2\pi]$ (the use of 6 symbols was found to be sufficient to differentiate the behaviors of L and F while maintaining a computationally feasible number of sequences to be sampled to compute information measures, see \cite{Basak2020,Basak2021}). For each set of initial conditions, the joint probability distribution of $p(x(t), y(t), y(t+1))$, from the time series, where $x(t)$ and $y(t)$ are the discretized forms of $\theta_{i}$, where $x$ and $y$ can be either $L$ or $F$. $p(x(t), y(t), y(t+1))$ is computed by counting the occurrences of each of the $6^3$ possible combinations of $(x(t), y(t), y(t+1))$ and dividing by the total length of the time-series minus 1, $2\times10^6-1$. Once the probability distributions are computed, $\mathcal{T}$ is computed by plugging them into the equation $\mathcal{T}= \sum_{y_{t+\tau},y_t,x_t}p(y_{t+\tau},y_{t},x_t)\log_2 \frac{p(y_{t+\tau}|y_t,x_t)}{p(y_{t+\tau}|y_t)}$~\cite{Schreiber2000}, where $\tau=1$. Likewise, $\mathcal{M}$ is computed using the formula $\mathcal{M}=\sum_{x_t,y_{t+\tau}}p(x_t,y_{t+\tau})\log_2\frac{p(x_t,y_{t+\tau})}{p(x_t)p(y_{t+\tau})}$.

\subsection{Details of the modified Vicsek Model}
\label{sec: vicsek_detail}
\begin{align}
\langle\bm{\theta}(t)\rangle_{R,\textbf{\textit{w}},\vec{\bm r}^{t}}
  = \arctan \left[\sum^\prime_j w_{ij} \sin \theta_j(t)
  \bigg/ \sum^\prime_j w_{ij}\cos\theta_j(t)\right]
  ~,\label{S10}
\end{align}
and $\sum^\prime_j$ sums over all $j$ satisfying $\vert
\vec{r}_i^{t}-\vec{r}_j^{t} \vert \leq R$. In this, $\boldsymbol{w}$ is a
nonnegative asymmetric matrix whose $w_{ij}$ element determines the interaction
strength that particle $i$ exerts on particle $j$. $w_{ij}>w_{ji}$ whenever
particle $i$ is a leader and particle $j$ is a follower in our setting. Positions at the initial time $t=1$ are chosen randomly from a uniform distribution within the box of length $L=10$, and orientations  are chosen randomly from a uniform distribution on the interval $[0, 2\pi)$. The interaction radius $R$ is set to $R=3$. Positions are updated using Eq. \ref{eqn: position}, and the orientations  $\theta_{\rm F}$ and $\theta_{\rm L}$  are updated using Eq. \ref{eqn: weight}. The weighted interaction term in Eq. \ref{eqn: weight} is computed by

\begin{align*}
\langle {\bm \theta}(t)\rangle_{R,\textbf{\textit{w}},\vec{\bm r}^{t}}
&=\tan^{-1} \left[  \frac{\sum_{j:\mid\vec{r}_i^t-\vec{r}_j^{t}\mid\le R}\biggl[w_{ii}\sin {\theta_i^t}+w_{ij}\sin {\theta_j^{t}}\biggr]}{\sum_{j:\mid\vec{r}_i^t-\vec{r}_j^{t}\mid\le R}\biggl[w_{ii}\cos {\theta_i^t}+w_{ij}\cos {\theta_j^{t}}\biggr]}  \right ],\nonumber \\
\end{align*}
the derivation of which can be found in the appendix of~\cite{Basak2020}.

\bibliographystyle{naturemag}

\end{document}